# SIP APIs for Voice and Video Communications on the Web

Carol Davids
Illinois Institute of Technology
Wheaton, IL
davids@iit.edu

Alan Johnston
Washington University
St. Louis, MO
alan.b.johnston@gmail.com

Kundan Singh
Intencity Cloud Tech
San Francisco, CA
kundan10@gmail.com

Henry Sinnreich
Unaffiliated
Richardson, TX
henry@sinnreich.net

Wilhelm Wimmreuter
SCCT, Germany
wilhelm@wimmreuter.de

## ABSTRACT

Existing standard protocols for the web and Internet telephony fail to deliver real-time interactive communication from within a web browser. In particular, the client-server web protocol over reliable TCP is not always suitable for end-to-end low latency media path needed for interactive voice and video communication. To solve this, we compare the available platform options using the existing technologies such as modifying the web programming language and protocol, using an existing web browser plugin, and a separate host resident application that the web browser can talk to. We argue that using a separate application as an *adaptor* is a promising short term as well as long-term strategy for voice and video communications on the web.

Our project aims at developing the open technology and sample implementations for web-based real-time voice and video communication applications. We describe the architecture of our project including (1) a RESTful web communication API over HTTP inspired by SIP message flows, (2) a web-friendly set of metadata for session description, and (3) an UDP-based end-to-end media path. All other telephony functions reside in the web application itself and/or in web feature servers. The adaptor approach allows us to easily add new voice and video codecs and NAT traversal technologies such as Host Identity Protocol. We want to make web-based communication accessible to millions of web developers, maximize the end user experience and security, and preserve the huge global investment in and experience from SIP systems while adhering to web standards and development tools as much as possible. We have created an open source prototype that allows you to freely use the conference application by directing a browser to the conference URL.

## Keywords
Real-time Web communication; browser API; voice and video; SIP API for Web communications.

## 1. INTRODUCTION

At present, rich Internet applications (RIA) on the web and real-time interactive applications such as voice-over-IP (VoIP) do not interact either all all, or not seamlessly because the protocols, programming language APIs, developer tools and communities are distinct for voice and web applications. Most Internet applications on the web use HTTP [2] as the only application protocol. At the same time, the global voice communications for both fixed and mobile telephony use the Session Initiation Protocol (SIP) [7] standards for interoperability, but have not produced any significant new applications other than emulating legacy telephony services.

We believe that the disconnect between RIA and VoIP is due to technical as well as non-technical reasons: (1) web developers and organizations work in a fast-paced quick turn-around easy-to-use application mind set, and do not want to entertain the complexity of SIP-family of standards, (2) web organizations want to own their content and customer interactions and hence prefer proprietary protocols over standards-based open systems, and also (3) critical programming primitives such as UDP transport, listening socket and native device access that are needed for VoIP are missing in present web languages and browsers. With the tremendous growth and innovation on the web in recent years, web developers started using browser plugins and server gateways to support communication within the limitations of a browser.

Hundreds of applications exist for audio/video communication. Some examples are standalone Skype, browser-based Gmail video chat, Flash Player based TinyChat and iPhone's Facetime. Note all of them use standards in their design and also have key parts proprietary. Even though the signaling and control technology behind these are drastically different, every real-time communication application tends to establish some form of end-to-end UDP media path, and falls back to relays if that fails. IETF standards exist to establish such media paths, end-to-end or via relays.

Research [10] has identified that only two protocols are required for web communication applications: (1) HTTP for signaling and control, basically data, including rendezvous, and (2) UDP for real-time media transport. All other application or telephony specific functions are not embedded in the network protocols but can reside in an application in the user client and/or in a web feature server. With this insight, we started our *voice and video on web* project with the main objective to develop the technology and sample implementations for web-based real-time communication [4]. Our project aims for web communication widgets to become as common on web pages as other components such as layout, buttons, images and multimedia players. We believe that the transformational benefits for both web applications and communications are (1) in enabling millions of web developers to include communications on their web pages and (2) in the seamless integration of web applications with communications having the potential of new, innovative applications.

More recently, standards bodies have identified two parts to web communications [8, 12]: (1) specific HTML/Javascript extensions to enable new elements for devices, codecs, and communication, and (2) on-the-wire protocol to enable end-to-end communication among browser instances. While standardization of these tasks in W3C and IETF may take a few years, we focus on pre-standard implementations using existing technologies and describe the short and long term benefits of our project.

Imagine a standard-compliant application that runs on user's machine independent of the browser, but allows any application including browser to establish real-time media-path. The browser can use existing HTTP to interact with this *adaptor* application.

The adaptor is not owned by a specific vendor, but is installed by the end-user. This avoids re-implementing the feature by every application developer who wants to do real-time communication. The main advantage of this approach is that it does not require changing the browser or HTML. It can easily add new voice or video codec or NAT traversal technology, and can be used by web and desktop applications alike. Our proposed API is inspired by the modern RESTful web services [6] known to web developers and uses the lessons learned from SIP systems, albeit in a web friendly manner.

The paper is organized as follows: Section 2 lists some differences of web communication with SIP systems. Section 3 compares available platform options, describes related work and lists the benefits of our approach. Section 4 describes the project architecture in detail using message flows, API description and preliminary implementation. Finally, we present conclusions and future work in Section 5.

## 2. DIFFERENCES WITH SIP SYSTEMS

In theory, an Internet SIP system follows the end-to-end principle of the Internet: keep the intelligence in the endpoint (or user agent) because the IP network and SIP proxies are transparent to applications in the endpoints. In practice however, a commercial SIP provider creates a closed walled garden using smart network elements (aka intermediaries) to prevent your SIP-capable device from directly using a third-party service without going through your provider's billable and "managed" services. Unlike telephony model, the web has evolved differently because the end-user is not tied to a specific web site. The difference with SIP systems conforming to the trapezoid mode is that interoperability between two web sites is usually a non-issue as all communicating parties are on the same web site. By keeping the signaling part outside the standard, we avoid the walled garden debate for web communication, and let the web site implement it in its own taste.

Typically a SIP-based user facing application is either a software rendition of a phone or a phone book to talk to your friends. There are other behind-the-scene components such as rendezvous server, application gateway or conference server. On the other hand, web communication is more immersive in what the end user is already doing on the web. For example, if you want to call a phone number listed on a web page, you want to click there to call instead of starting your SIP phone to make the call. This logical phone embedded in the web page is unlike a regular phone as it is meant to dial one number, and does not receive calls. Another example is if you are visiting a news page, and want to see who else is reading on the page and chat with them within the browser. You do not want to add them to your phone book or send them emails to invite them in a call. For example, in social networks, communication models may fundamentally differ from say a legacy business telephone call where a secretary forwards an incoming call to her manager. Thus, the web communication shifts the focus from telephony to immersive web communications within your browsing or collaboration experience.

The trust model and related message flows are also different. Typically, a user is already authenticated, e.g., using Google account when accessing its cloud applications such as email. A third-party application can re-use this user identity to provide services, e.g., many web sites now use other authentication services such as Facebook to connect and to authenticate a user for posting comments in its discussion forum. Similarly, a web communication application would use existing user identity provided by any third-party instead of asking the user to create a SIP account on every communication enabled web site. Thus, the web allows a multitude of authentication technologies ranging from simple passwords to ID cards or PKI services.

A caller in SIP typically invites a callee, joins a conference or is invited to a conference, before the session begins. In particular there is an explicit session invitation with offer/answer of session description. On the other hand, a web page can advertise its session parameters and any visitor can start the session by landing on that page. The invitation mechanism is outside the scope, and is usually done out of band via email, instant messaging or through other web pages.

For these reasons, we believe that many of the traditional SIP call flows are not quite relevant to web communications. However, some lessons from SIP are applied to web communication in our architecture.

## 3. PLATFORM OPTIONS

Since current browsers lack the full capability to support audio and video communication, we compare the available architecture choices and platform options as shown in Fig. 1. Our external application approach is shown in Fig. 1 (b)

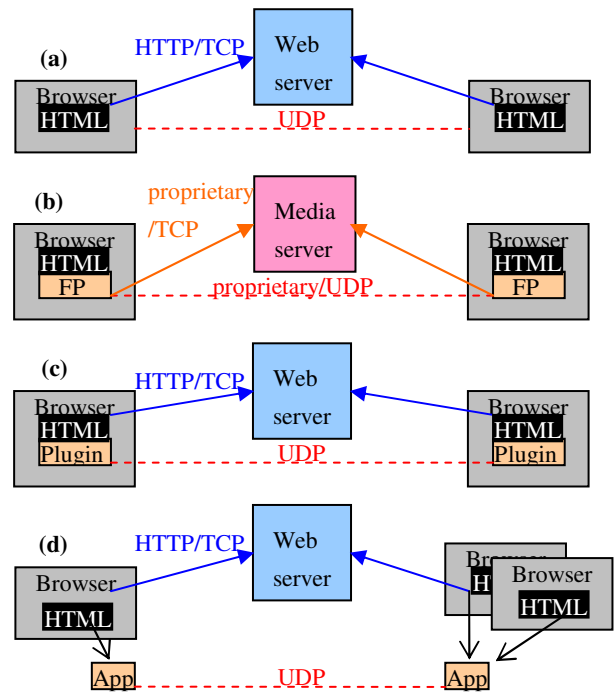

**Figure 1. Platform Options: (a) extend web protocol and language, (b) use Flash Player plugin, (c) use new browser plugin, or (d) use external application (App).**

### 3.1 Extend Web Protocol and Language

One approach is to improve the HTML/Javascript to support real-time communication [8, 12]. An existing browser such as Firefox can be modified to include the missing features. These features are exposed to the web application using new elements similar to how Ajax defines XMLHttpRequest or as HTML5 has the video element.

Depending on the granularity of the API, it can define high level registration and session objects, or low level camera, microphone and media connection objects as discussed next.

### 3.1.1 Implement SIP/RTP in Browser

Implementing SIP and related protocols in the browser enables existing applications to use the emerging standard SIP API such as to register, to make outbound calls, to receive and accept or reject an incoming call from SIP services. However, the SIP-family of protocols is very complex due to numerous extensions such as call transfer, conferencing and various other telephony services, and is described in thousands of pages of RFCs. We can consider only a subset of standards that is needed for a SIP user agent to turn your browser in to a programmable SIP user agent[11]. We consider that even with a small set of standards, the problem is too complex for all browser vendors to agree on a consistent API in HTML/Javascript.

### 3.1.2 Add New Codecs and Transport in Browser

Unlike the previous approach, in this approach the signaling part uses existing HTTP and websocket technology, but only the media and transport are added to the browser using a low level API. A websocket [13] allows converting an HTTP request connection to a persistent general purpose TCP connection for client-server data exchange. The application can define the data that goes on the TCP connection, e.g., using JavaScript object notation or XML. Web developers can thus pick any asynchronous custom protocols and data formats for rendezvous and session negotiation. The new API in the browser allows an application to use real-time codecs and capture devices of the local host and create end-to-end media path between two browser instances using NAT traversal techniques.

This approach is in-line with the evolution of web protocols as it uses HTTP as the only network application protocol and defines only a minimum set of new primitives to represent devices and connections, to connect a device with a connection, and to select preferences such as desired codecs. The resulting HTML/Javascript application is complex due to the low level API.

There are some open questions: (1) Should the minimum set of audio and video codecs be defined or be left to the browser vendor? (2) Should RTP be used for media transport between browsers or do we need another layer of multiplexing to reduce open bindings at the NAT? (3) Should interoperability with SIP systems be done in the browser or a separate gateway? (4) Should the connection be used for only the media path or also for other data communications, such as IM? (5) Should it enable interoperability between different web sites or leave it to web developer to configure cross domain authentication? (6) Should the end-user give permission to allow new connections or should it be controlled by cross domain policy of web sites? Hopefully, the new standards working groups in the IETF and W3C will be able to resolve the issues.

The main advantages of modifying web protocols are (1) no other dependency on external plugin or application besides the browser, (2) modifications can eventually be included in standards, and (3) numerous web developers can contribute to building applications.

The problems with this approach however are that users are generally reluctant to change their browser, even after new standards have emerged and hence getting ubiquitous user adoption may take a long time, dealing with device interfaces in a portable manner is a challenge, and device access and sharing across multiple instances of same browser or different browsers is not clear. In the past, incompatibility in HTML among browsers has been a nightmare for web developers, and extending HTML for yet another feature is bound to cause more interoperability problems. Browser vendors are sometimes not too keen to add a new feature, e.g., for business reasons, if it competes with the manufacturer's existing product or service. Two interoperability scenarios are significant: between browsers from different vendors running the same web page, and between two different web sites. The latter is tricky from security point of view if open standards are used.

## 3.2 Use Plugins such as Flash Player or Silverlight

Existing web-based video conferencing systems typically use a browser plugin such as Flash Player or Silverlight to work around the browser limitations [5]. The more popular Flash Player uses proprietary media transport protocols such as client-server RTMP over TCP and end-to-end RTMFP over UDP. Beyond just an audio/video player, it is a virtual machine to execute application code and provides secure and portable access to computer resources such as camera and microphone.

The main advantages of using Flash Player are: (1) ubiquitously available to almost everyone with a computer and an Internet connection, (2) browser agnostic implementation, (3) excellent developer tools for familiar web programming languages and a fast application development cycle, (4) provides integrated and rich web browsing experience, and (5) requires no additional installation for most users. Compared to other platform options described here, the Flash Player approach works with little effort because all the complexity is hidden in the plugin.

The main problems with browser plugins are: (1) while they supports outbound TCP connection, the cross domain restriction allows only closed, proprietary implementations of application protocols, e.g., SIP, (2) they lack general purpose UDP transport and listening sockets needed for VoIP, and (3) they do not give access to encoded audio and video data to the application hence one cannot build a standard compliant VoIP phone in the browser. While people have built gateways to translate between Flash Player and standard SIP/RTP systems, in general the closed nature of plugins means that the web developer depends on the plugin vendor, e.g., for echo cancellation, new codec, portability to new a device, and security updates.

## 3.3 Building a New Browser Plugin

Instead of using Flash Player, one can build a new browser plugin to perform media transport and processing. It works with or without an existing plugin, e.g., the need to delegate the media capture and playback to Flash Player. In that case you only need to implement missing pieces in the plugin, e.g., UDP transport, TCP listening socket and real-time codecs to be used by the Flash application. The main problems are portability across operating systems and browsers and user adoption of the new plugin. It also has limited flexibility once the plugin is deployed.

## 3.4 Use Separate Adaptor Application

We understand the limitations of a web browser and HTML, and do not "add" audio/video communications to it. Instead of improving the web browser using new protocols or plugins, we take a more general approach of a standalone application or service that runs on the user's host computer for real-time communications. Local browsers as well as other local applications can talk to this separate voice/video-on-web adaptor application using its HTTP-based API to enable real-time communication as shown in Fig. 1(d). Alternatively, a Flash application in the web page can use that API to control the adaptor. The user interface of the application is shown in the browser, but the actual video communication is done in the adaptor.

Using an adaptor application to fix existing software is not new, for example a P2P-SIP adaptor [9] running on a local machine or network can turn a client-server SIP phone to P2P-SIP. The adaptor approach is similar to the Google Mail plugin that enables video chat display within the browser using Flash Player but instead uses its own communication protocol in the external plugin process to enable end-to-end media path and standard voice/video codecs.

Unlike other approaches, where the application dies as soon as you close the web page, our adaptor is a persistent long-lived service. This has several advantages, e.g., we can pre-detect NAT and firewall configuration, pre-detect closest media relays for low latency, and/or build a distributed peer-to-peer network for scalability and robustness. The separate application can keep track of persistent communication state even when the user goes from one web page to another. Moreover, for NAT traversal, a host specific technology, e.g., Host Identity Protocol (HIP) [3] can be efficiently implemented only with our approach, whereas other approaches need to use per session solution, e.g., Interactive Connectivity Establishment (ICE). HIP also provides VPN-like security, IPv6 capability, mobility and multihoming. We believe that a general purpose NAT/firewall traversal solution is superior to application-specific one in the long term.

To address privacy and security concerns, the adaptor must directly ask permission from the end-user before initiating or accepting a connection or using devices. This is similar to how Flash Player asks the end-user for permission to capture from microphone or camera.

The main advantages of our approach are: (1) it has the flexibility of using any transport protocol including UDP, adding any new codecs or NAT traversal technologies, (2) one can use portable programming languages such as Python or Java to quickly build it, whereas modification to browser typically requires C/C++, (3) it is browser independent and hence easier to implement, (4) one can use it together with a Flash application for portable device access to further simplify the implementation.

**Table 1. Platform options: (a) modify web protocols, (b) use browser plugin, (c) use separate application**

| Properties | (a) | (b) | (c) |
|---|---|---|---|
| With existing technology | No | Yes | Yes |
| Emerging standard protocol | Yes | No | Yes |
| Allows walled garden | Difficult | Easy | Difficult |
| Requires new install | No | Maybe | Yes |
| App dies on page close | Yes | Yes | No |
| Re-use web security means | Yes | Yes | No |

The main problems are: (1) it requires yet another installation by the end user and this potentially hampers wide adoption, (2) security and access control requires careful design to prevent unauthorized access and leaking of private information in the adaptor to the web page or to other users on a multi user system, (3) it is difficult to re-use authentication mechanisms already provided by standard web-browsers for media path.

In summary, this approach benefits from being an independent application with less restriction as well as uses the simplicity, portability and flexibility of the web platform. Table 1 presents the summary.

## 4. ARCHITECTURE

Among the available platform options, the separate application raises some deployment concerns but seems to work well in enabling web communications in light of existing tools and constraints. The block diagram in Fig. 1(d) shows three components: client, server and a separate adaptor application that runs on the same computer. The web server provides signaling using HTTP and asynchronous bidirectional channels such as websocket. The client (web browser) runs HTML, a Javascript or Flash application to display the front-end widget. Note the communication application has two components: (1) the business logic, for example the video-telephone state machine and (2) the user interface which is critical for a pleasing experience. The client also communicates with the adaptor using an authenticated API. In this section we describe the client-server communication for signaling, the client-side communication widgets, and the functions and API of the adaptor. The availability of free code and lack of license restrictions may facilitate its large scale adoption.

### 4.1 The Signaling API

The client-server communication is called the signaling API as it is used for rendezvous or "match making" among web users for communication.

*4.1.1 Requirements*

1) It should support long-lived connection so that asynchronous events can be delivered when needed. In practice, several options exist for this such as Bidirectional Stream over Synchronous HTTP (BOSH), Comet-style Javascript programming, new websocket and Google App Engine's channel API backed by XMPP.

2) It should be based on RESTful architecture in-line with modern scalable web services [6]. A resource is identified using an URL, and allows operations such as create, read, update or delete using HTTP methods POST, GET, PUT or DELETE, respectively.

3) It should allow publish/subscribe style communication, so that one can subscribe to a resource URL and get notified when the resource changes. For example, a user subscribes to his voice mail resource to get notified when someone adds a new item there.

This needs two new methods to subscribe and notify on the web URL, similar to the SIP event framework. To work with existing systems that do not support these additional methods, one can use an URL parameter, e.g., command=subscribe, to indicate a subscribe request. A subscription creates a long-lived connection to receive asynchronous notifications.

The actual API is dependent on the application, but here we give an idea using an example of two-party voice communication system. Similar to a SIP registrar, the web server keeps track of online users, and facilitates rendezvous. Consider an example, in which a web site keeps track of all the users visiting that site. It displays a list of online users. When a visitor clicks on an online user's name, the visitor's web page sends call invitation to the user, and instructs the adaptors of the two parties to establish an end-to-end voice path. This works well, but is not RESTful.

*4.1.2 Example of a RESTful API*

Let us define a RESTful API for this client-server communication. The web application server provides two core resources, /login and /call, to represent a list of currently logged in users and list of active calls. The client uses standard HTTP with XML or JSON data format, e.g., using Ajax in the web page, to access and manipulate these resources as shown in Fig.2. Dotted lines indicate regular HTTP request-response, whereas solid lines are for subscribe-notify messages over long-lived HTTP.

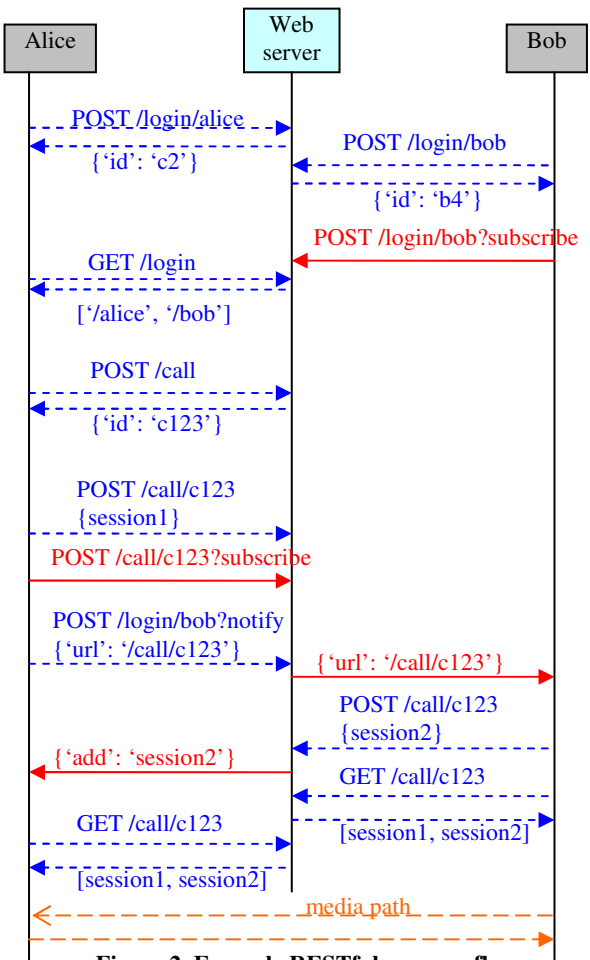

**Figure 2. Example RESTful message flow**

*Login*: The SIP user registration is mapped to the /login/{email} resource. For example, Alice does POST /login/alice@example.net with a request body containing the additional contact information including listening media transport candidates to register. The response contains an identifier, c2, for this contact resource which can now be accessed at /login/alice@example.net/c2. Later, she can use DELETE on it to unregister this contact or PUT to update it. The actual representation of the login contact resource can be in XML, JSON or plain text and is application dependent. One can combine the presence update including rich presence with the registration method. Existing data formats defined in various SIP presence specifications can be re-used. Clearly the login update requires appropriate authentication, but standard web authentication works well here. Doing a GET on /login gives list of current online users. Additionally, URL parameters such as offset=20&limit=10 allow pagination of result. To know if a particular user is online or not, do GET /login/{email}. A registering user also subscribes to her contact resource to receive notifications sent to her.

*Call*: The call logic is split in to two parts: conference resource and invitation. The conference resource is /call/{call-id}, where a client can POST /call to create a new call identifier, or POST /call/c123 to join an existing call, c123. The conference resource represents the list of participants. Again, the data formats defined in the SIP centralized conferencing can be re-used.

Call invitation is optional in the API as it can be done via other means, e.g., sending a web URL via email, instant messaging or another web page. Within the API, call invitation can be done by using the notify request on the callee's login resource, e.g., using POST /login/bob@example.net?command=notify with the request body containing the invitation attributes such as conference resource, time of invitation, return notification data. We need notification for both call invitation as well as cancellation. When the callee accepts the call, he also joins the same conference resource.

Each participant subscribes to the conference resource, so that he can get notification about membership change. Each conference participant resource has session parameters such as media stream URL for centralized conference, or transport data and media capabilities of his client for end-to-end media path. The session parameter includes media transport addresses as well as supported and preferred media codecs. Instead of using outdated SDP, we use web friendly XML or JSON to format the application specific session parameters. Once it learns the session parameters of the other participant, it can initiate end-to-end media path.

Thus, a RESTful interface for web communication signaling is feasible using existing web protocols and tools. The goal is not to replace SIP, but to provide a new mechanism that allows web-centric applications to use communication services and to allow building such easy to use application servers. Other RESTful resources are also possible, e.g., /user/{email} can represents a signed up user, and has sub-resources for profile data, voice mail, contact list, etc.

The signaling API is invoked by the client application or widget. Although, Fig. 2 shows a lot of messages, it ensures simplicity of the REST API. In practice, scalable web servers can handle many such requests without perceived signaling latency. For example, a browser typically downloads many additional files (script, style, image, etc.) when displaying a single web page today.

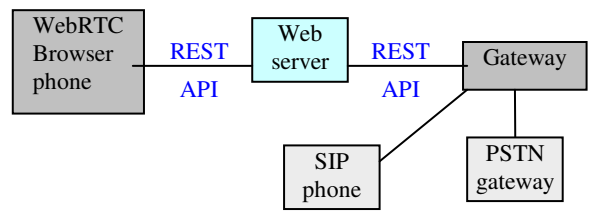

**Figure 3. Gateway translating between REST and SIP**

### 4.1.3 Other use cases for RESTful APIs

In addition to web browsers, the signaling API can be used by a gateway to interoperate between HTTP APIs on one side, and SIP or PSTN on the other side as shown in Fig. 3. It provides interworking between web browser and legacy communication systems based on SIP.

## 4.2 Communication Widgets

From a web developer's point of view, it should be easy to add new communication primitives to a web page similar to other elements such as a text input or multimedia player. Following are some example communication widgets.

1) A click-to-call label or text-input, which when clicked initiates a VoIP call to the target browser, phone number or SIP address-of-record (AoR). The state of the label or text-input updates to indicate the call progress, and also allows the user to end the call. The text-input can store call history using a drop-down combo box. In a web page this appears as a clickable text, image or edit box.

2) A contact list object that displays zero or more contacts similar to that on existing instant messengers. It displays user presence by getting the data on asynchronous communication channel. It can be used to display regular contact list as well as conference members list. For example, a web site that allows its visitors to chat with each other may bind the contact list object with the active visitor list resource.

3) A web phone object that allows a web user to make or receive voice and video call within the browser. The web developer configures certain attributes of the object such as server addresses. The video displays are bound to local or remote media streams, and laid-out in picture-in-picture or tile mode. It includes an asynchronous communication channel to receive call events.

4) A web conference object that allows a visitor to join an existing voice, video and/or text conference. Unlike a web phone, this does not receive call invitations, but represents a joined conference. Additionally, it implements optional conference controls by the owner or moderator, or the web site itself.

The list is obviously not complete, but gives an idea about the high level elements that web developers expect. A widget uses the API of the adaptor and talks to the web server. In particular, a widget includes HTML/Javascript code which can use native communication support in the browser if available, or fall back to Flash Player plugin for media. A web developer can create new widgets to support more web applications using the adaptor API.

## 4.3 The Adaptor API

As discussed before, the separate adaptor application implements the media and transport for web communications. The client widget connects to the adaptor on the local host over a long-lived HTTP connection to issue commands and to receive notifications. If the widget cannot connect, it assumes that the adaptor is not running and prompts the user to download and install it. Once installed, it runs as a service in the background. Some widgets are distributed with the adaptor itself so that web pages can access them immediately, while others must be downloaded from third-party web sites.

After connecting, the adaptor authenticates the client application. It then uses the transport and media classes to implement end-to-end media path for voice and video communication. When the widget creates a new object, e.g., UDP transport, it gets a unique object identifier within the authorized scope of its authentication key. The HTML web application can use this object identifier as a regular Javascript object and invoke methods on it or install callbacks to receive notifications. The widget internally translates the method to an RPC over HTTP to the adaptor so that the latter can actually perform those API functions.

The adaptor exposes several object-oriented transport and media related classes to the widget. We use web friendly data format such as XML or JSON (Javascript Object Notation) for our command and notification. For any sensitive method, the adaptor prompts the end user using a native dialog box to approve the API request. The widget should gracefully handle any request denial.

### 4.3.1 Application Authentication

Each application that connects to the adaptor needs approval from the end user. On first connection, the adaptor issues a time-bound secure token to the application so that a subsequent connection by the same application does not need approval. This allows for the same application to be distributed across different web pages or even multiple browser instances, e.g., one browser window for each participant video in a conference. The user can ask to always allow a particular application, in which case a permanent secure token is generated by the adaptor and stored on the web server by the client application. For better security, the adaptor may require web site's certificate-based identity from the web application.

### 4.3.2 Transport Classes

The adaptor implements several transport related classes. These transport objects enable high level application protocols such as vanilla SIP or low level transport connections such as an ICE session. A UDP transport implemented using the datagram socket can bind to any ephemeral port (higher than 1024) to receive packets as a server or send packets as a client. The adaptor prompts the user for approval before binding and sending/receiving for first time from/to a target IP address. A TCP transport with optional secure attribute for TLS and using an outbound stream socket connection is enough for a TCP bound media path, especially with UDP-blocking firewalls. ICE based transport combines multiple UDP and TCP transports in to a single logical object to hide the complexity of approval and data handling. The API allows initiating the various phases of ICE and allow sending of data after the connection setup is successful. Similarly, the RTP transport contains two UDP transports, for RTP and RTCP, as a single logical object. It should be possible to merge an RTP, ICE and HIP transports, so that multiple functions are included.

### 4.3.3 Media Classes

The four basic classes, Microphone, Speaker, Camera and Display, represent the corresponding audio and video functions and are implemented natively in the adaptor. The API includes the attributes such as sound volume and codec name. The video Display object due to rendering and size requirements is particularly difficult to implement in a browser without using a plugin. Initially, we plan to keep the API of these components similar to that in the Flash Player so that we can leverage it for capture and display if needed. The client can connect these device objects to each other or to a transport. Unlike Flash Player, the client can also get access to the encoded audio and video data if needed. The adaptor prompts for approval if the media data is sent to the client.

## 4.4 Preliminary Implementation

Our preliminary implementation is available as open source and open client-server API [4]. Fig. 4 shows two screen captures from a web conference among the authors in San Francisco, Chicago, Dallas and Munich. The demonstration uses Google Chrome browser and integrated Flash Player 10.3 with built-in echo cancellation. Performance depends as of this writing strongly on the type of machines used.

The client software in the browser has in principle several components: the protocol machinery emulating SIP or some other signaling data exchange, the real time media transport for audio and video such as RTP or proprietary, the audio codec and echo control, the video codec, the business (call) logic, and the user interface. These components can be quite independent and each of them can be implemented using either standards or proprietary approaches, such as in Flash or Silverlight.

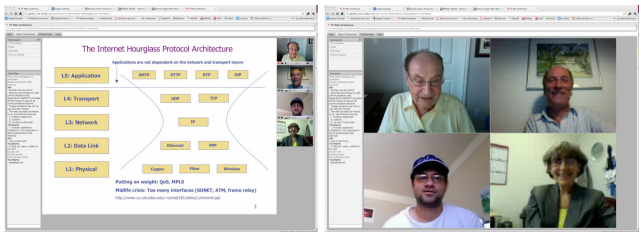

**Figure 4. Screen captures of (a) presentation and (b) voice/video/text chat modes during a web conference**

The current implementation has two separate parts: (1) the signaling protocol machinery, business logic and the graphical user interface are written using standard HTML, CSS and JavaScript code, (2) the media part is using the Flash Player plugin as an intermediate solution until we implement the separate application. At present, the Flash Player solves several hard problems in a bundled way: real-time end-to-end media transport, NAT traversal, audio and video encoding, echo control and end-user control over access permission to the audio/videos resources of the machine. As standards emerge, some or all of these parts may be adapted in the various conference modules.

## 5. CONCLUSIONS AND FUTURE WORK

We propose a voice and video communication architecture on the web and compare it with other alternatives. Using a separate adaptor application has several advantages compared to modifying the web protocol and language or using the Flash Player plugin. In particular, the architecture is platform independent, can be easily implemented, and is flexible enough to accommodate new codecs, application protocols (e.g., SIP/RTP) or NAT/firewall traversal techniques in the long term. For instance, with the emergence of CDN, one can easily use media over HTTP for many application scenarios using the adaptor approach [1]. In the short term, our approach is easy to implement because of browser independence, and use of existing web protocols and tools. One major challenge is to dynamically find the best quality settings that perform satisfactorily across various platforms and networks.

Security and privacy concerns need to be carefully considered to avoid misuse of user's resources, leaking private conversation data, or overwhelming the user or adaptor with too many requests. Analyzing the more trustworthy flavors of authentication such as open-ID, info-cards and PKI can seed deployment and overcome known trust issues for principals and relying parties. As a result we can avoid or reduce current authentication and trust issues of SIP by re-using web authentication technology. This will lower the barrier to use strong authentication for SIP calls and allows a healthy re-use of available web-based identity and authentication mechanisms for real time voice and video communication services. Another challenge is that we require a new installation, which may slow down user adoption.

A future task is to re-factor the implementation and various APIs so that a module can be easily replaced once a standard emerges, e.g., the real-time media transport. By standards, we mean here both de jure and some dominant market based standards. For commercial quality products, where user experience matters more than standards, a complete plugin or proprietary technology based implementation should also be possible.

Our current and future work involves implementing a prototype of the separate application using client-side adaptor in C/C++. We plan to build a few more widgets using end-to-end media path and optional fall back to client-server tunneling. The RESTful client-server API and client-side widgets are useful for other platform options as well, e.g., with the upcoming WebRTC standards.

## 6. REFERENCES

[1] Amirante, A., et al. NTRULO: A tunneling architecture for multimedia conferencing over IP. *NEW2AN'10*, St. Petersburg, Russia. pp 460-472. Aug. 2010

[2] Fielding, R., et al. Hypertext transfer protocol – HTTP/1.1. RFC 2616. *IETF*. Jun.1999

[3] Komu, M., et al. Basic HIP extensions for traversal of NAT. RFC 5770. *IETF*. Apr.2010

[4] Project: voice and video on web. *Illinois Institute of Technology*. https://sites.google.com/site/vvowproject/

[5] Project: Flash based audio and video communication. http://code.google.com/p/flash-videoio/

[6] Richardson, L., Ruby, S. RESTful Web Services. *O'Reilly*. May 2007. ISBN 978-0-596-52926-0

[7] Rosenberg, J., Schulzrinne, H., et al. SIP: session initiation protocol. RFC 3261. *IETF*. Jun.2002

[8] RTC-Web IETF working charter proposal. Mar.2011. http://rtc-web.alvestrand.com/ietf-activity

[9] Singh, K., Schulzrinne, H. SIPpeer: a SIP-based P2P Internet telephony client adaptor. Implementation Report. *Columbia University*. New York, NY. 2004

[10] Sinnreich, H., Johnston, A. SIP APIs for communications on the web. IETF Internet draft. "work in progress". Jun 2010

[11] Sinnreich, H., Johnston, A., Shim, E., Singh, K. Simple SIP usage scenario for applications in the endpoints. RFC 5638. *IETF*. Sep.2009

[12] Web real-time communications working group charter. *W3C*. Dec.2010. http://www.w3.org/2010/12/webrtc-charter.html

[13] Fette, I., The websocket protocol, IETF Internet draft, "work in progress". Jun 2011.